\documentclass[cameraready]{Interspeech}
\usepackage{pifont}
\usepackage{multirow}

\newcommand{\xmark}{\ding{55}} 

\title{MuVAP: Multimodal Multiparty Voice Activity Projection for \\Turn-taking Prediction in the Wild}

\author[affiliation={1},orcid=0009-0002-2946-0054, correspondingauthor]{Haotian}{Qi}
\author[affiliation={1},orcid=0000-0002-8579-1790]{Gabriel}{Skantze}

\address{
    $^1$ Department of Speech Music and Hearing, KTH
Stockholm, Sweden
}

\email{haotianq@kth.se, skantze@kth.se}


\usepackage{comment}

\begin{document}

\maketitle

\begin{abstract}
    Current multiparty turn-taking models often rely on complex microphone arrays or multi-camera setups, limiting their applicability in human-robot interaction scenarios. We introduce MuVAP, a causal multimodal framework that extends Voice Activity Projection by grounding acoustic predictions in face tracks, enabling speaker-aware turn-taking predictions from a monaural audio stream and a single camera view. To address the combinatorial complexity of modeling multiple speakers, we propose Role-Relative Projection, which maps any N-speaker interaction onto a fixed current versus next floor-holder state. Because existing audiovisual datasets contain disruptive editing cuts that break causal tracking, we introduce the Audio-Visual Conversation Corpus, a 31-hour dataset of unedited, single-camera multiparty conversations. Evaluations demonstrate that MuVAP outperforms strong baselines on Shift-Hold and next-speaker prediction tasks across two- and three-speaker settings.
\end{abstract}

\section{Introduction}

Turn-taking is a fundamental aspect of conversation. Since people cannot easily speak and listen simultaneously, they must coordinate their turns through a complex exchange of cues \cite{duncan1972some,sacks:74,Duncan1974OnTS}. For example, a syntactically or semantically incomplete phrase may signal a turn hold, whereas a complete phrase may be turn-yielding \cite{ford_thompson:96}. A filled pause is a strong cue for turn-holding \cite{clark100117, Ball1975}. When syntax and semantics are ambiguous, prosody and visual cues such as lip motion and facial dynamics can also be informative \cite{Duncan1977FacetofaceIR}. These cues allow humans to take turns with very small gaps (around 200 ms \cite{stivers2009universals}), while avoiding large overlaps or interruptions \cite{levinson_torreira:15}.

\begin{figure}[t]
  \includegraphics[width=\linewidth]{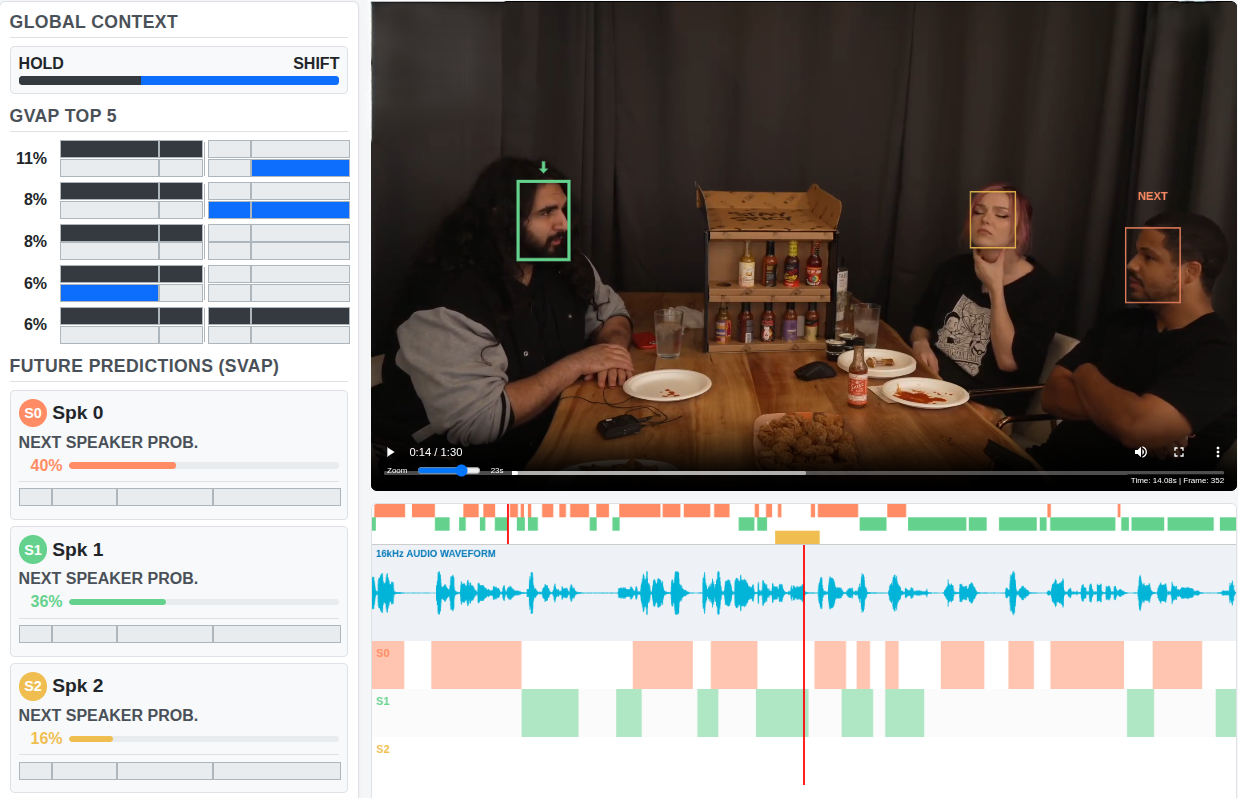}
  \caption[Visualization demo]{Visualization demo\footnotemark[1] of multiparty turn-taking prediction. The top panel shows the video stream with tracking of speakers S0 (red), S1 (green), and S2 (yellow), with the single channel audio stream below. The bottom panel shows the ground truth speech activity timelines, with the red vertical line marking the current time step $t$, leading to a turn shift from S1 to S0. The left panel shows the GlobalVAP (GVAP) shift/hold predictions along with individual SpeakerVAP (SVAP) predictions.}
  \label{fig:example}
\end{figure}

\footnotetext[1]{\url{https://github.com/Haotian-Qi/MuVAP}}

Traditional conversational systems have mostly relied on silence thresholds to trigger system responses \cite{Skantze2021}. This reactive approach leads to unnatural delays and interruptions, motivating a shift towards continuous, predictive turn-taking. Notably, \textbf{Voice Activity Projection (VAP)} \cite{vap} utilizes acoustic features and formulates turn-taking as a continuous projection of speech activity into the near future (e.g., 2 seconds). By forecasting the joint activity of speakers, VAP can predict complex phenomena, including turn shifts, backchannels, and interruptions, in a zero-shot fashion \cite{vap} or by training a linear probe on the learned embeddings \cite{inoue2025yeah}. 

Recent work has shown how the VAP model can be integrated into conversational systems, including human-robot interaction (HRI), to improve turn-taking and backchannel predictions \cite{inoue2025yeah, skantze2025applying}. However, for HRI applications, traditional VAP suffers from two critical limitations. First, it is predominantly designed for two-party interactions, whereas HRI frequently involves multiparty settings. Second, it often relies exclusively on the speech channel. In multiparty dialogue, the visual channel is indispensable, since the model must determine not only if a turn is ending, but \textit{who} among the group is the next speaker.

We propose \textbf{MuVAP} (Multimodal Multiparty Voice Activity Projection), a framework that anchors voice activity projections to faces detected in the video stream. Unlike prior approaches that rely on separate audio channels or assume fixed dyadic interactions, MuVAP operates in unconstrained multiparty settings without requiring channel separation. This design enables the model to jointly address two complementary tasks. First, it predicts the temporal onset of forthcoming speech (\textit{when}) by leveraging audiovisual and prosodic cues. Second, it infers the identity of the current and likely next speaker (\textit{who}) by grounding its predictions in the visible participants within the scene.

To address the combinatorial complexity inherent in multiparty interactions, we introduce \textbf{Role-Relative Projection} (Figure~\ref{fig:objective}). Rather than modeling all possible speaker–listener permutations, we adopt a role-centric abstraction inspired by human conversational behavior: interlocutors primarily monitor the current floor-holder and the most probable next speaker. By projecting group dynamics onto these two relative roles, MuVAP attains social scalability -- enabling a single, fixed architecture to generalize to an arbitrary number of participants without retraining or architectural modification.

\begin{figure}[t]
  \includegraphics[width=\linewidth]{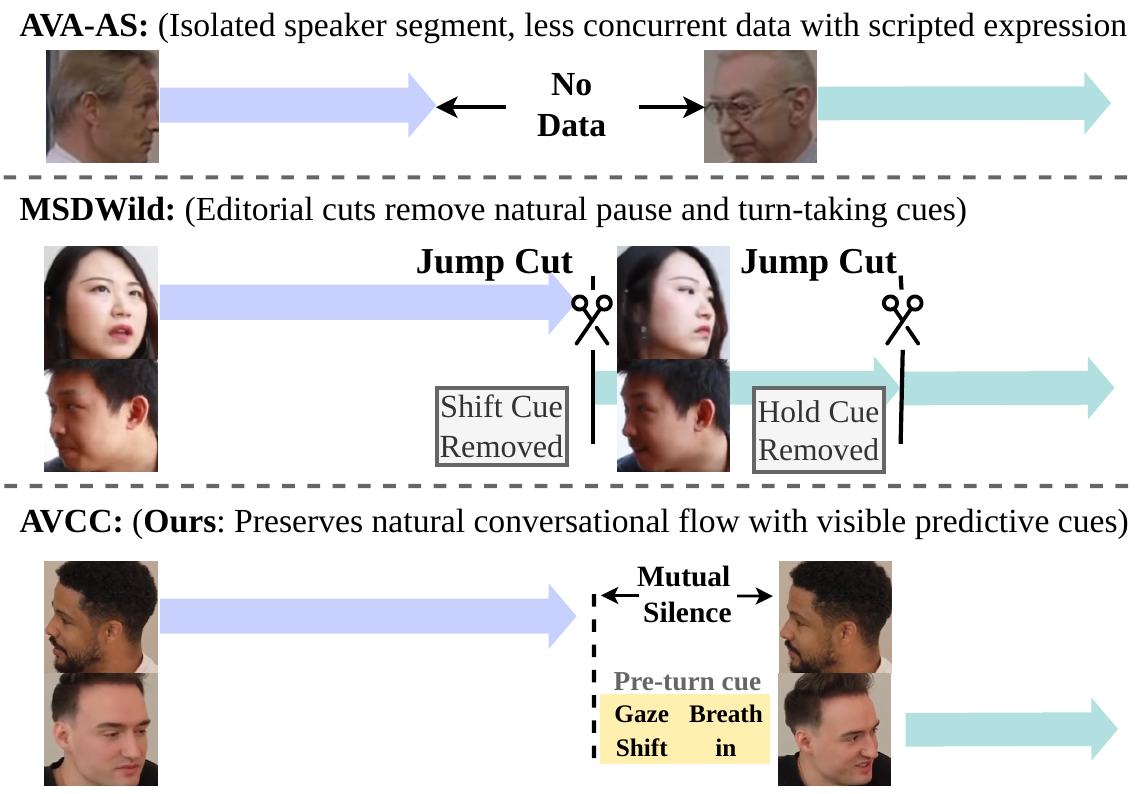}
  \caption {Comparison of dataset continuity and structure. AVA-ActiveSpeaker (Top) is limited to many isolated speaker segments rather than full scene, often relying on scripted cues that lack natural turn-taking dynamics. MSDWild (Middle) captures unconstrained scenes but suffers from editorial and jump cuts that disrupt the visual timeline needed for prediction. In contrast, AVCC (Bottom) preserves the unedited, continuous flow of conversation. 
  }
  \label{fig:continuity}
\end{figure}

In contrast to previous studies \cite{lee2023multimodal, elmers2025triadic} that rely on data from laboratory settings, our work leverages unconstrained, real-world video. 
While some existing datasets used for Active Speaker Detection (ASD), such as MSDWild \cite{msdwild}, cover more naturalistic settings, they typically contain \textit{editing gaps}, which make them unsuitable for directly training a causally grounded turn-taking framework like MuVAP.

To address this, we introduce the \textbf{Audio-Visual Conversation Corpus} (AVCC) by collecting and annotating approximately 31 hours of unedited videos of multiparty dialogue from the web, specifically filtered for static single-camera perspectives to preserve genuine social dynamics. Furthermore, to leverage existing high-resource data, we adopt a modular training strategy that utilizes multiple datasets as shown in Table \ref{tab:dataset_comparison}: (a) a large-scale telephone corpus (1960 h) to learn acoustic turn-taking signals, (b) standard ASD datasets (140 h) to pre-train the ASD module, and (c) the AVCC dataset to optimize the temporal turn-taking dynamics on multimodal, continuous, unedited sequences of conversation.

While previous studies have examined multiparty turn-taking prediction, ASD, and multimodal interaction modeling, explicit real-time next speaker identity forecasting in multiparty settings under a strictly single-view, single-channel setting remains largely unexplored. We address this gap by proposing a causal multimodal model that continuously anticipates which participant will take the conversational floor, without exploiting spatial audio cues, microphone arrays, or multi-view visual geometry.

\begin{table*}[t!]
\centering

\begin{minipage}[t]{0.72\textwidth}
\centering
\begin{tabular}[t]{l c c c l l c}
\hline
\textbf{Dataset} & \textbf{Modality}& \textbf{Duration}& \textbf{Source}& \textbf{Lang.} & \textbf{Unedited}& \textbf{Module} \\
\hline
Fisher & Audio& ~1958h7m& Telephone & English & \checkmark & VAP\\
AVA-AS & AV& ~38h5m& Movie & Multi  & \xmark & ASD\\
MSDWild & AV& ~80h18m& Vlogs/Wild & Multi & \xmark& ASD\\
WASD & AV& ~30.0h& Wild & Multi & \xmark& ASD\\
\textbf{AVCC (Ours)} & AV& ~30h49m & Wild & English & \checkmark & MuVAP\\
\hline
\end{tabular}
\caption{Comparison of datasets used in this work. 
AV = Audio-visual. Unedited means continuous, uncut conversations essential for long-term turn-taking modeling.}
\label{tab:dataset_comparison}
\end{minipage}\hfill
\begin{minipage}[t]{0.24\textwidth}
\centering
\begin{tabular}[t]{l c}
\hline
\textbf{AVCC Subset} & \textbf{Duration} \\
\hline
2-speaker & 17h31m \\
3-speaker & 13h21m \\
\hline
Training & 22h28m \\
Validation & 8h24m \\
\hline
\textbf{Total} & \textbf{30h52m} \\
\hline
\end{tabular}
\caption{Distribution of the collected AVCC dataset by speaker count and partition.}
\label{tab:avcc_stats}
\end{minipage}

\end{table*}

\section{Related Work} 

\subsection{Active Speaker Detection vs. Turn-Taking}

We distinguish our work from multimodal Active Speaker Detection (ASD), which aims to identify the active speaker among multiple faces in a video stream. Current ASD architectures, such as TalkNet \cite{talknet} and LoCoNet \cite{wang2024loconet}, achieve impressive precision on the AVA-ActiveSpeaker (AVA-AS) benchmark \cite{roth2020ava} by correlating lip motion with audio streams. While sometimes conflated with turn-taking, ASD differs strictly in scope and temporality. 

Standard ASD models focus on identifying the \textbf{current speaker state}. However, they often overlook the underlying conversational context, such as whether a speaker is mid-utterance, providing a brief backchannel, or yielding the floor during silence. In contrast, MuVAP models \textbf{interaction dynamics}. This approach captures the subtle, predictive signals that reveal a participant's intent to either maintain, yield, or seize the floor. We also enforce strict causality throughout the architecture. While many ASD benchmarks improve detection accuracy by looking at future frames, MuVAP is designed for the reality of live interaction. Every prediction is made using only the history available at that point.

\subsection{Voice Activity Projection (VAP)}

Voice Activity Projection (VAP) \cite{vap} is a self-supervised framework designed to predict future conversational patterns directly from raw acoustic inputs. Unlike traditional turn-taking approaches that rely on reactive silence thresholds or specific event labels like pauses or overlaps, VAP formulates turn-taking as a continuous forecasting problem. It projects the joint voice activity of two speakers into a near-future window (typically 2000 ms). This enables Next Speaker Prediction (NSP) tasks within strict dyadic settings, where a \textit{Shift} indicates that the listener will take the turn.

To make this objective tractable, the future window is grouped into activity bins, and the joint activity bins are mapped to a finite codebook of discrete labels. By optimizing a cross-entropy objective over these future states, VAP implicitly learns to recognize complex coordination cues, such as prosodic shifts, backchannels, and filled pauses, without requiring manual event annotations \cite{prosody, vap}.

However, the standard VAP formulation faces a critical bottleneck in multiparty settings due to a combinatorial explosion, as seen in Figure~\ref{fig:objective}. Because VAP models the joint state of all speakers simultaneously, the label space grows exponentially with the number of speakers ($N$). While recent extensions have successfully adapted VAP for fixed triadic settings \cite{elmers2025triadic}, this joint-modeling approach cannot scale to arbitrary or dynamic group sizes; it also suffers from coarse label resolution. The codebook would require retraining for every possible value of $N$, which makes standard VAP unsuitable for dynamic settings where the number of participants fluctuates.

\subsection{Multimodal Next Speaker Prediction}

Next Speaker Prediction (NSP) is a classification task in which the system must predict which speaker, within a multiparty conversation, will take the next turn. Traditional frameworks rely on explicit pre-utterance indicators like gaze transition patterns \cite{ishii2013predicting, heo2025gaze, onishi2023multimodal}, head pose dynamics \cite{lee2023multimodal}, and mouth movement \cite{ishii2019prediction} to identify floor acquisition. Speakers use these visual signals alongside prosodic contours to navigate turn-taking \cite{petukhova2009s}. While sentence-level NSP \cite{lee2024computational} and event-based visual frameworks \cite{roddy2018multimodal, russell2025visual} achieve high classification accuracy, they treat turn-taking as isolated events, missing the continuous alignment of acoustic and facial dynamics.

A major hurdle for existing NSP models is their reliance on controlled or high-resource environments. Many state-of-the-art models \cite{cheng2025multi, lee2025enhancing, heo2025gaze, russell2025visual} require a multi-channel microphone array or a dedicated camera for every speaker to successfully isolate signals. Such hardware dependencies are impractical for real-world Human-Robot Interaction (HRI) because they cannot easily adapt to unconstrained settings. Conversely, existing single-channel monaural systems that lack this hardware support often suffer from identity drift, especially when speakers have similar vocal timbres.

MuVAP distinguishes itself by operating under a strict single-camera and single-audio-stream constraint. Unlike previous works that depend on multi-channel inputs, MuVAP avoids explicit pose extraction and treats the visual modality primarily as a separation anchor. By maintaining spatially distinct visual tracks within a single video stream, we anchor the global acoustic history to the correct speaker identity. This architecture allows the model to predict short periods of future speech activity for each speaker simultaneously. Consequently, the system can identify the next speaker at turn transitions using only the "in the wild" data available from a single monaural source and one camera view.

\section{The AVCC Dataset}

Currently, publicly available ASD datasets \cite{kopuklu2021asdnet,msdwild} suffer from domain gaps that limit their utility for turn-taking prediction in interactive settings (such as HRI). As illustrated in Figure~\ref{fig:continuity}, these sources contain editing artifacts like jump cuts that disrupt the temporal flow of conversation. This prevents models from learning the true causal history required to predict turn-taking. Conversely, egocentric datasets \cite{grauman2022ego4d} introduce viewpoint bias in which the camera wearer's actions influence sensory input. Similarly, while multiparty datasets like AMI \cite{kraaij2005ami} capture continuous dialogue, they rely on overhead cameras and dedicated lenses for each participant. This distributed array captures a view that is fundamentally disconnected from the singular visual stream a human or robot relies on.

The \textbf{Audio-Visual Conversation Corpus}\footnotemark[1] (AVCC) dataset was collected according to three primary criteria to ensure suitability for causal turn-taking modeling. First, we selected only continuous, unedited sequences to preserve the natural temporal flow of conversation. This allows the model to observe the full duration of mutual silences and pre-turn behaviors. Second, we restricted the data to a static third-party perspective. This constraint prevents the model from relying on camera motion or editorial cues, forcing it to focus on participant-level signals. Third, we prioritized spontaneous social interactions to capture realistic phenomena like overlaps, interruptions, and hesitation cues among familiar friends. We collected the data from YouTube and Twitch livestreams of unscripted multiparty conversations. These formats were chosen because they consist of unscripted dialogue where natural turn-taking and conversational fillers are frequent. 

In total, the dataset comprises 30 hours and 52 minutes of recordings, spanning both two-speaker and three-speaker interaction settings. The distribution of the number of speakers, as well as training/validation splits are shown in Table~\ref{tab:avcc_stats}. Table \ref{tab:dynamics} also shows some descriptive turn-taking statistics for the dataset, compared to the Fisher telephone dataset. As can be seen, AVCC exhibits wider inter-quartile ranges, confirming that unconstrained interaction in the wild is less predictable than dyadic telephone speech, even in 2-speaker settings.

\label{sec:data}
\begin{table}[t]
  \caption{Conversational dynamics of Fisher (Telephone) vs. AVCC (Multimodal) with 2 or 3 speakers. Numbers are median seconds with inter-quartile ranges, reported for Gaps (silence between turns), Pauses (silence within turns) and Overlaps (between turns). }
  \label{tab:dynamics}
  \centering
  \begin{tabular}{l c c c}
    \toprule
    \textbf{Metric} & \textbf{Fisher}& \textbf{AVCC (2)}& \textbf{AVCC (3)}\\
    \midrule
    Gaps & 0.32 & 0.39 & 0.43 \\
     & [0.16--0.68] & [0.15--0.88] & [0.15--1.00] \\
    \midrule
    Pauses & 0.40 & 0.50 & 0.61 \\
     & [0.24--0.76] & [0.27--1.01] & [0.36--1.13] \\
    \midrule
    Overlaps & 0.56 & 0.44 & 0.50 \\
     & [0.24--1.48] & [0.17--1.21] & [0.19--1.23] \\
    \bottomrule
  \end{tabular}
\end{table}
 
We utilized the InsightFace \cite{deng2021masked} framework with RetinaFace \cite{Deng2020CVPR} backbone for automated face detection and identity tracking. We localized faces using SCRFD \cite{guo2105sample} with a confidence threshold of 0.62. To eliminate background artifacts, we applied a relative area filter that dropped any detections smaller than 10\% of the largest face within a given frame. To initialize stable speaker identities for the dataset, we sampled 50 random frames from each video to establish a set of global anchor embeddings. These anchors, in combination with frame-by-frame embeddings generated by ArcFace \cite{deng2018arcface}, were used to maintain temporal continuity.

We calculated a joint cost matrix by combining the cosine distance of these embeddings with the Euclidean distance of bounding box centroids. Speaker assignment was resolved using the Hungarian algorithm \cite{kuhn1955hungarian}. This approach ensures that speaker tracks remain stable across all segments of a video. We performed a manual audit of the resulting tracks, checking for identity swaps or missed detections by identifying sudden changes in the distance of bounding box centroids.
The synchronized audio and resulting face crops were processed by our pre-trained ASD model, as shown in Figure \ref{fig:architecture}, with a VAD objective to generate initial voice activity labels for each speaker. All annotations were then manually refined using the VIA annotator to ensure high-fidelity ground truth labels, similar to those shown in Figure~\ref{fig:example}.

\begin{figure}[t]
  \includegraphics[width=\linewidth]{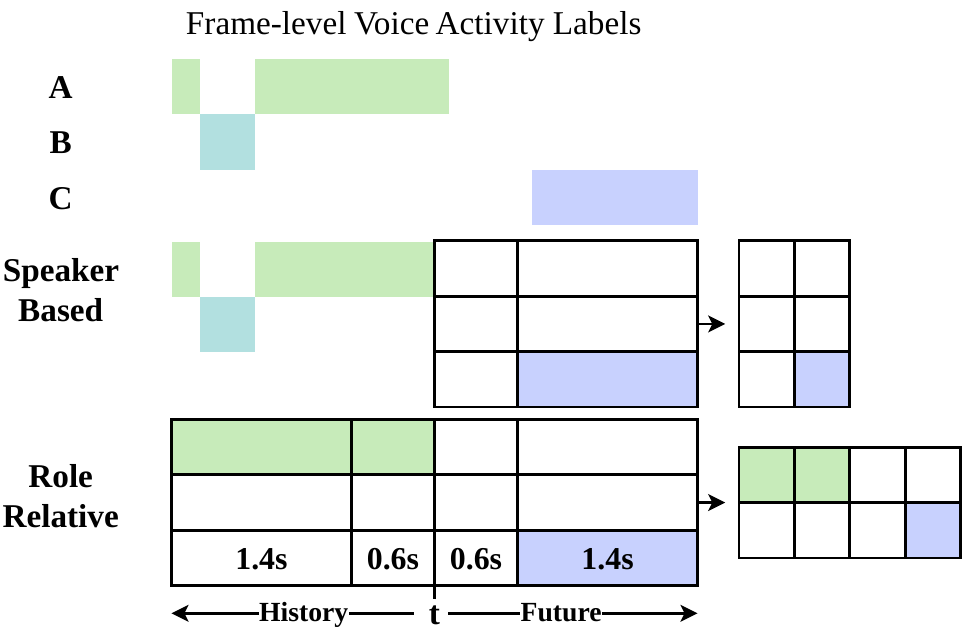}
  \caption {The difference between speaker based projection \cite{elmers2025triadic} and \textbf{(Ours)} Role-Relative Projection for 3-speaker conversations.}
  \label{fig:objective}
\end{figure}

\section{Model}
We propose \textbf{MuVAP} (Multimodal Multiparty Voice Activity Projection), a causal framework designed to forecast turn-taking dynamics for an arbitrary number of speakers. Given a single-channel audio waveform $\mathcal{A} \in \mathbb{R}^T$ and a set of face tracks $\mathcal{V} = \{v_1, \dots, v_N\}$ for $N$ detected speakers, the model predicts two probability distributions at each time step $t$:

\begin{itemize}
    \item GlobalVAP \textbf{(GVAP)}: The joint turn-taking state of the ``current floor holder'' versus the ``next floor holder''.
    \item SpeakerVAP \textbf{(SVAP)}: A speaker-specific projection for each tracked face for next speaker predictions.
\end{itemize}

\begin{figure*}[t]
\centering
  \includegraphics[width=\linewidth]{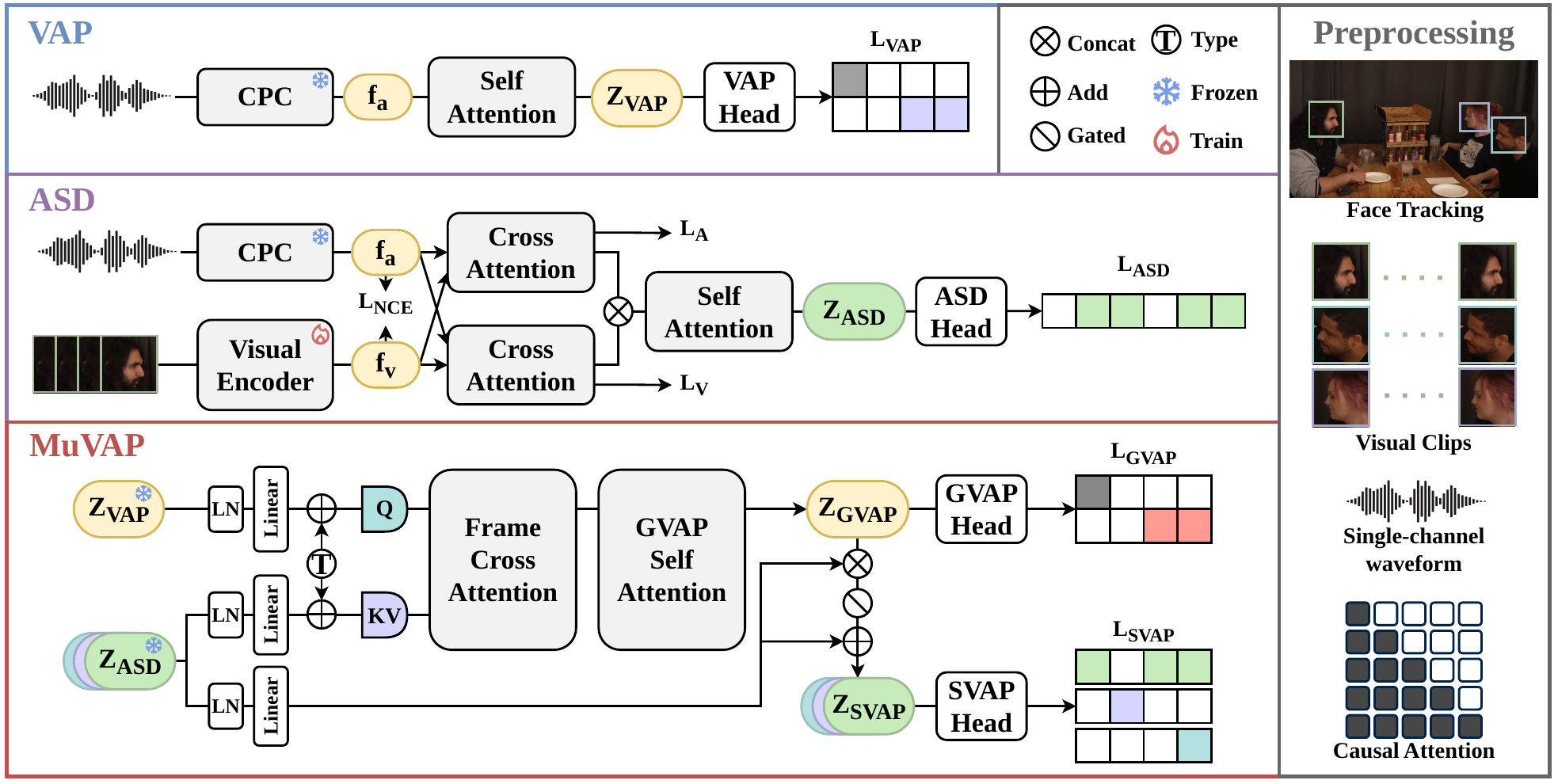}
  \caption {Modular architecture of the MuVAP model. The embeddings from the VAP and ASD modules ($Z_{\text{VAP}}$ and $Z_{\text{ASD}}$) are used as input to the main module (right), which makes both GlobalVAP (GVAP) and SpeakerVAP (SVAP) predictions.}
  \label{fig:architecture}
\end{figure*}

The overall architecture and the different sub-modules are shown in Figure~\ref{fig:architecture}.

\subsection{VAP Backbone}

The goal of the VAP module is to extract prosodic and linguistic cues from the audio waveform features \cite{prosody}. Following \cite{vap}, we utilize a contrastive predictive coding (CPC) encoder to map raw audio to dense representations, followed by a causal downsampling convolution (100Hz $\rightarrow$ 25Hz) to align with the video frame rate. These audio features then pass to a Transformer \cite{transformer} block with ALiBi positional embeddings and causal masking.

For the original VAP model \cite{vap}, the prediction target at each incremental step was defined by a window of 2 seconds containing the future voice activity for both speakers. In that setting, the window was discretized into 8 separate bins (4 for each speaker), with each bin corresponding to time bins of increasing duration in seconds $[0.2, 0.4, 0.6, 0.8]$s. Each bin was assigned a value of 1 if more than half of its frames contain active speech from that speaker. This 2 $\times$ 4 discretization yields an 8-bit binary vector, corresponding to 256 unique classes.

\subsubsection{Role-Relative Projection}

\label{sec:role-relative}

As noted earlier, the original VAP objective does not scale well to multiparty settings, as the number of unique classes grows exponentially with the number of speakers (e.g., $2^{4\times N}$, where $N$ denotes the number of speakers). Moreover, when only a single mono audio channel is available, it is not possible to anchor predictions to individual speakers. The original VAP model \cite{vap} also used a mono channel, but it made use of individual voice activity labels as input to anchor the predictions. Our model only uses a mono channel, without any such labels as input. 

To resolve this, we introduce \textbf{Role-Relative Projection}, where the objective is to predict future voice activity \textit{relative} to past voice activity, within certain time bins, by letting the classes jointly represent both future and past time bins, as shown in Figure~\ref{fig:objective}. This reduces the variable $N$-speaker state to a fixed ``Current vs. Next'' pairwise representation. This reduces temporal resolution in bin durations, but is based on the simplifying assumption that most turn-taking typically involves two out of the $N$ speakers at any given point in time. 

For each time step $t$, we discretize voice activity into a window of four bins: two history bins and two future bins as $[1.4, 0.6, 0.6, 1.4]$s duration for each speaker, as shown in Figure~\ref{fig:objective}. We then determine the projection labels via a two-stage ranking:

\begin{enumerate}
    \item Current Holder ($S_{\text{curr}}$): We rank all $N$ speakers by their total activity in the history bins. The speaker with the highest activity is designated as the current/past floor holder.
    \item Next Holder ($S_{\text{next}}$): We rank the remaining $N-1$ speakers by their activity in the future bins. The highest-ranked remaining speaker is designated as the primary next speaker.
\end{enumerate}

This reduction transforms the complex multiparty dynamic into a fixed pairwise state $\{S_{\text{curr}}, S_{\text{next}}\}$. Following the standard VAP encoding method \cite{vap}, we extract the binary activity for this pair (2 roles, $\times$ 4 bins) to form an 8-bit vector. 
While an 8-bit vector theoretically require $2^{2\times4}=256$ states, 
we make the codebook invariant to the ordering of the two
patterns by treating
$(S_{\text{curr}},S_{\text{next}})$ and
$(S_{\text{next}},S_{\text{curr}})$ as the
same state, resulting in 136 unique states. Thus,

\begin{align*}
    c([0011,1100]) = c([1100,0011]) \\ 
    c([0111,1000]) = c([1000,0111]) \\ 
\end{align*}



The VAP module is trained to minimize the cross-entropy loss between the predicted probability distribution $\hat{y}_t$ and the ground truth label $y_t$ over the sequence length $T$ from the latent embedding $Z_{\text{VAP}}$:

\begin{equation}
    L_{\text{VAP}} = - \frac{1}{T} \sum_{t=1}^{T} \log P(\hat{y}_t = y_t \mid Z_{\text{VAP}})
\end{equation}

\subsection{Causal ASD Backbone}
To ground predictions in visual tracks and distinguish audio between participants, we require robust multimodal embeddings that focus on speech separation based on visual grounding. We pre-train a dedicated ASD module with three datasets. We adopt a modified TalkNet architecture \cite{talknet}, replacing non-causal temporal convolutions with causal ones, increasing the dilation to [1, 2, 4, 8, 16], and using a pre-trained CPC audio encoder (the same used for the VAP module). 

Unlike the traditional binary ASD objective that focuses on current frames, we use a similar bin configuration as the VAP module, with history bins of [0.8, 0.6, 0.4, 0.2]s and future bins [0.2, 0.4]s. We use shorter future bins with the goal of still capturing certain backchannel expressions within edited video datasets, which results in a 6-bin prediction of $y_{\text{asd}}$. Note that since this module makes predictions for each speaker independently, the activity of the different bins is not predicted jointly, but as independent binary predictions. 

The goal is to minimize the binary cross-entropy loss for the 6-bin target using the fused embedding $Z_{\text{ASD}}$. Additionally, we utilize the side streams ($L_\textbf{a}$ and $L_\textbf{v}$) for auxiliary tasks. Despite sharing the underlying encoders, these branches are constrained to predict the binary voice activity ($y_{t}$) for the current frame $t$ as an auxiliary loss. We also utilize the TalkNCE \cite{jung2024talknce} loss, which performs contrastive loss on the features $f_a$ and $f_v$ to pull the positive pairs closer, denoted as $L_{\text{NCE}} $, to boost the active speakers' visual representations with the CPC audio embeddings. The total loss is a weighted sum, where all coefficients follow the configuration in \cite{jung2024talknce}:

\begin{equation}
L_{\text{total}} = L_{\text{ASD}} + 0.4 (L_{\text{a}} + L_{\text{v}}) + 0.3 L_{\text{NCE}}
\end{equation}

\subsection{MuVAP: Multimodal \& Multiparty}

We treat multiparty turn-taking as a hierarchical coordination problem. The GlobalVAP (GVAP) acts as a ``social conductor'', monitoring the global pulse of the conversation to predict \textit{when} a transition is imminent. This global context then guides the SpeakerVAP (SVAP), which focuses on individual visual tracks to resolve \textit{which} specific participant will take the next turn.

To fuse different modalities, the model takes global audio embeddings $Z_{\text{VAP}}$ and a set of $N$ individual speaker embeddings, $\{Z^{1}_{\text{ASD}}, \dots, Z^{N}_{\text{ASD}}\}$ from ASD modules. These inputs are processed through independent LayerNorm and linear projection layers to map them into a shared 256-dimensional space to match the VAP embeddings, ensuring the representations are aligned before the attention blocks.

We then use the $Z_{\text{VAP}}$ as the \textit{Query} and $Z^{N}_{\text{ASD}}$ as \textit{Key} and \textit{Value}, passing them through a frame transformer to allow the model to attend to all speaker tokens within the same temporal frame. We then pass the output through another temporal transformer to capture temporal dynamics. This yields the $Z_{\text{GVAP}}$ embeddings, which capture the global turn-taking dynamics for multiparty settings. 

We pass the input $Z^{N}_{\text{ASD}}$ embeddings through separate LayerNorm and projection layers, then use gated addition with the $Z_{\text{GVAP}}$ to let the GVAP dynamically modulate the speaker features. The modulated features are fed into the SVAP prediction head where the target consists of independent binary predictions of future bin activity with durations [0.2, 0.4, 0.6, 0.8]s, which matches the original VAP bin resolution. The training objective is as follows:
\[L_{\text{MuVAP}} = L_{\text{GVAP}} + \frac{1}{N} \sum_{n=1}^{N} L_{\text{SVAP}}^{(n)}\]

\section{Implementation}

Our pipeline utilizes five distinct datasets: Fisher (Parts 1 \& 2), MSDWild, WASD, AVA-ActiveSpeaker, and our newly introduced AVCC dataset. Each dataset is mapped to specific modules to facilitate multi-stage training as shown in Table \ref{tab:dataset_comparison}.

All modules were optimized using a Cosine Annealing learning rate scheduler with a linear warm-up phase covering the first 10\% of total training steps.  The learning rate reached a peak of $1 \times 10^{-3}$ and decayed to $1 \times 10^{-4}$ over five epochs, utilizing a Weight Decay of $0.01$. Experiments were conducted on a single NVIDIA A100 (40GB) GPU. The model consists of 27.7 million total parameters. This is distributed across three modules: the ASD module (20.7 M), the VAP module (4.9 M), and the MuVAP module (2.1 M).

For preprocessing, face crops were extracted based on tracked coordinates and missing frames were padded with a grayscale value of 127. Video streams were standardized to 25 fps, and audio was sampled at 16 kHz in mono.

\subsection{VAP Backbone}
\label{sec:vap-backbone}
The VAP module was trained on the Fisher Corpus of spoken dialogue over telephone. We partitioned the data by session ID: sessions divisible by 16 were reserved for validation, while the remainder formed the training set. For the logistic regression probe, groups \{064, 080\} were held out as \textit{Test} set and the rest were used as the \textit{Fit} set to train on the linear probe. Audio samples were segmented into 30-second windows with 10-second overlaps. The temporal Transformer comprises $L=4$ layers, $h=4$ attention heads, and a hidden dimension of $d_{model}=768$. We use ALiBi positional encoding and set the dropout to $0.1$.

\subsection{ASD Backbone}
The ASD backbone was trained on a composite of MSDWild, WASD, and the AVA-ActiveSpeaker training set. We modified the standard TalkNet architecture by replacing all non-causal temporal layers with causal convolutions. We also increased the dilation to $[1, 2, 4, 8, 16]$ for more visual history context. The cross-attention mechanism has a hidden dimension of 768, while self-attention blocks were expanded to 1024 dimensions. Due to high memory requirements, we employed a batch size of 4 during training.

\subsection{MuVAP}
The complete MuVAP model was trained on the AVCC dataset using a batch size of 16. During this phase, the VAP and ASD backbones were frozen to serve as robust feature extractors. Audio ($Z_{\text{VAP}}$) and visual ($Z_{\text{ASD}}$) embeddings were projected into a shared 256-dimensional latent space through independent Linear-LayerNorm blocks before being processed by the Frame transformer and Temporal transformer. Both transformer blocks have a feed-forward expansion factor of 4, with 4 attention heads and 1 layer.

\begin{figure}[t]
  \includegraphics[width=\linewidth]{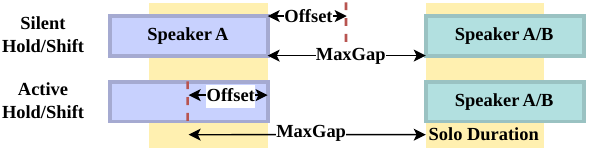}
  \caption {Visualization of the Active and Silent prediction sampling methods. The yellow region indicates the required solo-speaker duration (1000 ms) and offset (100 ms) used to prevent label leakage. The maximum gap duration is 3000 ms for both event types. Active events are derived from Silent events by shifting the prediction point into the ongoing speech segment while maintaining a fixed 100 ms offset from the speech endpoint. }
  \label{fig:tasks}
\end{figure}

\section{Downstream tasks}
\label{sec:downstream}

In the spirit of the original VAP model \cite{vap}, MuVAP is trained to learn generic embeddings capturing turn-taking dynamics. To measure the usefulness of these learned embeddings, we define a set of downstream tasks for evaluation. 

\subsection{Shift-Hold vs. Next Speaker Prediction} 
\label{sec:downstream-sil-shift-hold}

We first distinguish between two types of tasks: The first task is \textbf{Shift-Hold Prediction}, where the model has to determine whether the turn is about to \textit{Shift}, or whether the current speaker will \textit{Hold} the turn. This binary task does not require specific speaker identification and relies solely on raw role-relative VAP/GVAP predictions. Thus, it can be done on both multimodal (AVCC) and unimodal (Fisher) scenarios. This downstream task is performed using a linear probe (a Logistic Regression model) trained on top of the $Z_{\text{GVAP}}$ embeddings. To prevent data leakage, we strictly separate the data used for the probe. We extract the event prediction points from the AVCC Train partition to fit the linear probe, and we evaluate its performance exclusively on the AVCC Validation partition. This split was based on the number of speakers and scenario settings in each video to ensure a balanced distribution of 2-speaker and 3-speaker configurations across all subsets, as seen in Table \ref{tab:data_dist}.

The second task is \textbf{Next Speaker Prediction} (NSP), where the model has to determine \textit{who} (among the $N$ speakers) will be the next speaker, in moments of mutual silence. These decision points are the same as those for the Shift-Hold Prediction task, but the decision is not relative to the previous speaker. Note that the set of potential next speakers includes the current speaker (equivalent to a Hold). To solve this task, the raw VAP/GVAP embeddings are not sufficient; the model also has to make use of the individual SVAP predictions. Note that the NSP task in a 2-speaker scenario is not exactly the same as Shift-Hold Prediction, since the model has to successfully anchor the prediction in the set of visible speakers, which is not strictly necessary in the Shift-Hold Prediction task.

Additionally, we propose a GVAP-conditioned filtering strategy to ensure individual speaker assignments remain consistent with the projected global state. By utilizing the GVAP prediction as a conductor prior, we logically constrain the candidate set. If the GVAP predicts a \textit{Shift}, the identified previous speaker is eliminated from the pool of potential candidates. If a \textit{Hold} is predicted, the next speaker is automatically assigned to the previous floor holder. This hierarchical coupling forces the speaker-level predictions to align with the overall rhythm of the group interaction.

To identify the previous floor holder, we aggregate SVAP predictions over a 1-second rolling window, effectively using the 0.2s future-bin predictions as a temporally shifted proxy for recent speech activity, and assign the speaker with the highest cumulative probability. To calculate the likelihood of a specific participant taking the floor, we sum the probabilities of the final two future bins from their respective SVAP head. By comparing these scores across all participants, the model identifies the most probable next speaker.

\subsection{Silent vs. Active Prediction} 

We then distinguish between two different ways of determining \textit{when} to make the prediction: First, we identify points of \textit{mutual silence} (\textbf{Silent Prediction}), where a speaker has just stopped speaking and no other participant is speaking. These prediction points are illustrated in the top part of Figure \ref{fig:tasks}. Prediction points are sampled at a +100 ms offset from the end-of-speech. We only include pauses/gaps $\le$ 3 s that are preceded by at least 1 s of single-speaker speech.

A more proactive version of this task is to predict upcoming turn dynamics while a speaker is still active. Unlike silent prediction, which occurs after a vocal offset, we define \textbf{Active Prediction} by sampling points within the ongoing speech of the current floor holder. These points are derived directly from the previously identified silence events by moving the prediction point earlier, into the active speech region. To ensure the model has sufficient causal history, we also shift the solo speaker requirement backwards to ensure that the prediction point is preceded by at least 1s of uninterrupted clean speech from a single participant.

For \textbf{Active-Shift} events, the prediction point is placed within the vocalization of an interlocutor whose turn results in a floor transition. In contrast, \textbf{Active-Hold} events are sampled from turns that precede a within-turn pause where the current speaker continues to hold the floor. By anchoring both classes to a fixed offset of 100ms before the speech ends, the task measures the model's capacity to decode terminal signals such as specific prosodic contours or stable facial dynamics that indicate turn-yielding or turn-holding intent. This setup evaluates if the model can anticipate an upcoming transition point before the actual onset of mutual silence. The distribution of labels for the training and validation of events are shown in Table \ref{tab:data_dist}.

\begin{table}[t]
  \caption{Split of the AVCC dataset Train and validation partition, used for the Logistic Regression probe. Samples are categorized by Shift-Hold mode (Active/Silent) and the number of speakers present (2 or 3).}
  \label{tab:data_dist}
  \centering
  \resizebox{\columnwidth}{!}{
  \begin{tabular}{llcccc}
    \toprule
    \textbf{Split} & \textbf{Mode} & \textbf{Total} & \textbf{2/3-Spk}& \textbf{HOLD} & \textbf{SHIFT} \\
    \midrule
    \multirow{2}{*}{Train}& Active & 17688& 10216/7472& 7900& 9788\\
                           & Silent& 17901& 10829/7072& 9952& 7949\\
    \midrule
    \multirow{2}{*}{Val}& Active & 5891& 3274/2617& 2631& 3260\\
                         & Silent& 6205& 3725/2480& 3623& 2582\\
    \bottomrule
  \end{tabular}%
  }
\end{table}

\section{Results}

We evaluate Shift-Hold Prediction using Macro-F1 (due to class imbalance), and Next Speaker Prediction (NSP) using accuracy. Results are reported as mean ± 95\% confidence intervals over ten runs using random seeds 42–51.

We compare MuVAP against two baselines: \textbf{Majority Class} for the binary Shift-Hold task and \textbf{Random} for the multi-class NSP task. To quantify the benefit of our proposed fusion strategy, we also implement a No-Fusion (\textbf{MLP}) baseline. This model processes VAP and ASD embeddings separately through individual LayerNorm and projection layers. These layers reduce the input dimension from 512 to 256. We apply a GELU activation and 0.1 dropout before passing to the NSP and Shift-Hold prediction heads, respectively.

\subsection{Role-Relative Projection}

Since we introduce a new VAP training objective in this paper -- the Role-Relative Projection (see Section \ref{sec:role-relative}) -- we first want to compare this with the speaker based projection in the original VAP model in a two-speaker scenario where we do have two audio channels, namely the Fisher corpus. We train two models based on these two objectives: One with mono audio and Role-Relative Projection, and one with stereo audio and speaker based projection. We then use the $Z_{\text{VAP}}$ embeddings to train a logistic regression probe for the Silent Shift-Hold Prediction task (as defined in Section \ref{sec:downstream-sil-shift-hold}) on the Fisher \textit{Fit} set and evaluate on the \textit{test} set (as defined in Section \ref{sec:vap-backbone}). 

The results are shown in Table \ref{tab:vap_fisher}. The stereo model has better performance than the mono model, which is expected, given that stereo channels provide explicit speaker attribution, which makes it easier to identify speaker shifts and resolve overlapping speech. In the mono setting (which is a requirement for our model), the model has to figure out speaker switches based on speaker characteristics in the speech signal. In light of this, we think that the drop in performance (about 2\%) is quite modest, and validates the viability of the Role-Relative Projection to be used in our model. 

\begin{table}[th]
  \caption{Performance comparison between speaker based and our Role-Relative VAP training objectives on the Fisher dataset.}
  \label{tab:vap_fisher}
  \setlength{\tabcolsep}{4pt}
  \centering
  \begin{tabular}{ll}
    \toprule
    \textbf{Objective} & \textbf{Macro-F1}\\
    \midrule
    Majority class         & .451\\
    Speaker based (stereo) & .799 \\
    Role-Relative (mono)   & .778 \\ 
    \bottomrule
  \end{tabular}
\end{table}

\subsection{Shift-Hold Prediction}

As described in Section \ref{sec:downstream} above, both the Silent and Active Shift-Hold prediction tasks are evaluated with a logistic regression probe trained on the $Z_{\text{GVAP}}$ embeddings. As baselines, we also use the unimodal $Z_{\text{VAP}}$ embeddings from the VAP model trained on Fisher (see box 1 in Figure \ref{fig:architecture}).

The results for the Silent and Active Shift-Hold Predictions are shown in Table \ref{tab:shift_results}. Our model outperforms the baselines across all tasks, indicating that visual information helps for Shift-Hold Prediction and that the gated fusion of the MuVAP model contributes. Results are similar between 2 and 3 speaker settings, even for the VAP model, indicating that Shift-Hold Prediction using Role-Relative Projection works well in both these settings, since individual turn shifts typically involve just two of the participants. While the overall performance in mutual silence is lower than in dyadic telephone settings (Table \ref{tab:vap_fisher}), these results demonstrate that MuVAP effectively handles the increased complexity of multiparty interactions. The performance is somewhat lower in the Active tasks, which is expected, but still indicates that turn shift forecasting is possible. 

\subsection{Next Speaker Prediction}

The results for NSP are shown in Table \ref{tab:nsp_result}. Again, our model outperforms the baseline MLP model across the different tasks and settings. We also show how the results can be further improved by not just selecting the speaker with the highest future SVAP predictions, but by also conditioning the prediction using the GVAP predictions to infer turn holds, in combination with SVAP predictions of the previous speaker (+GVAP). However, these predictions of the previous speaker are not perfect, as can be seen in Table \ref{tab:prev_result}. We therefore also show the potential performance if the previous speaker detection was perfect (+GVAP+GT), as an upper bound of the performance. 

As expected, for NSP, 3-speaker settings are harder than 2-speaker settings, since a turn shift might lead to any of the other two partners taking the turn. In many cases, if the current speaker does not select the next speaker, it is up to the other speakers to decide who would like to take the turn, so-called \textit{self-selection} \cite{sacks:74}, which is very hard, if not impossible, to predict.

\begin{table}[t]
  \caption{Performance on the Silent and Active Shift-Hold Prediction tasks on the AVCC dataset. Results for the MLP and our model are reported with 95\% confidence intervals across 10 seeds.}
  \label{tab:shift_results}
  \centering
  \begin{tabular}{l l c c}
    \toprule
    & \textbf{Model} & \textbf{2 Spk (F1)} & \textbf{3 Spk (F1)}\\
    \midrule
    \multirow{4}{*}{\rotatebox{90}{\textit{Silent}}}
    & Majority class & .367 & .351 \\
    & VAP            & .672 & .655 \\
    & MLP            & .650\scriptsize$\pm$.003 & .654\scriptsize$\pm$.003 \\
    & MuVAP          & .696\scriptsize$\pm$.003 & .670\scriptsize$\pm$.002 \\
    \midrule
    \multirow{4}{*}{\rotatebox{90}{\textit{Active}}}
    & Majority class & .346 & .367 \\
    & VAP            & .622 & .634 \\
    & MLP            & .610\scriptsize$\pm$.003 & .635\scriptsize$\pm$.002 \\
    & MuVAP          & .641\scriptsize$\pm$.005 & .652\scriptsize$\pm$.002 \\
    \bottomrule
  \end{tabular}
\end{table}

\begin{table}[t]
  \caption{Results (accuracy) for the Silent and Active Next Speaker Prediction (NSP) task on the AVCC dataset, with Random and MLP baselines, depending on speaker counts, with 95\% confidence intervals across 10 seeds. (+GVAP = conditioning on the GVAP and previous speaker predictions. +GVAP+GT = conditioning on GVAP predictions and ground truth previous speaker labels.)}
  \label{tab:nsp_result}
  \centering
  \begin{tabular}{l l c c}
    \toprule
    & \textbf{Model} & \textbf{2 Spk (acc)} & \textbf{3 Spk (acc)}\\
    \midrule
    & Random & .500 & .333 \\
    \midrule
    \multirow{5}{*}{\rotatebox{90}{\footnotesize \textit{Silent}}}& 
        MLP & .617\scriptsize$\pm$.002 & .464\scriptsize$\pm$.002 \\
        & MuVAP & .637\scriptsize$\pm$.003 & .477\scriptsize$\pm$.001\\
        \cmidrule{2-4}
        & MuVAP (+GVAP) & .666\scriptsize$\pm$.003 & .508\scriptsize$\pm$.003\\
        & MuVAP (+GVAP+GT) & .702\scriptsize$\pm$.003 & .547\scriptsize$\pm$.003\\
    \midrule
    \multirow{5}{*}{\rotatebox{90}{\footnotesize \textit{Active}}}& 
        MLP & .543\scriptsize$\pm$.001 & .429\scriptsize$\pm$.001 \\
        & MuVAP & .560\scriptsize$\pm$.002 & .441\scriptsize$\pm$.002\\
        \cmidrule{2-4}
        & MuVAP (+GVAP) & .605\scriptsize$\pm$.005 & .483\scriptsize$\pm$.003\\
        & MuVAP (+GVAP+GT) & .652\scriptsize$\pm$.003 & .516\scriptsize$\pm$.002\\
    \bottomrule
  \end{tabular}
\end{table}

\begin{table}[t]
  \caption{Results (accuracy) for inferring the previous speaker, based on SVAP-based predictions, with 95\% confidence intervals across 10 seeds.}
  \label{tab:prev_result}
  \centering
  \begin{tabular}{l l c c}
    \toprule
    & \textbf{Model} & \textbf{2 Spk (acc)} & \textbf{3 Spk (acc)}\\
    \midrule& Random& .500& .333\\
    \textit{Silent}& SVAP& .837\scriptsize$\pm$.001& .760\scriptsize$\pm$.001\\
    \textit{Active}& SVAP& .821\scriptsize$\pm$.001& .742\scriptsize$\pm$.001\\
    \bottomrule
  \end{tabular}
\end{table}

\section{Discussion}


The results reveal a clear division of labor between the modalities in our model. Acoustic, linguistic, and prosodic cues provide the global rhythm of coordination, while the visual channel acts as an essential separation anchor during competitive overlap and active speech. Rather than relying on explicit behavioral cues, our ASD modules ground the acoustic signal to individual visual tracks. The performance gain over unimodal baselines in Active Shift-Hold prediction shows that visual grounding helps disentangle overlapping acoustic signals and assign future activity to the correct participant. While this structural grounding proves effective, previous research demonstrates that explicit behavioral signals like gaze and head pose strongly govern turn allocation \cite{lee2023multimodal, heo2025gaze, onishi2023multimodal, russell2025visual}. Integrating these specific visual features on top of our separation tracks provides a clear path for future performance gains.


Role-Relative Projection can be seen as mimicking human social attention, to some extent. Multiparty turn-taking operates through a highly local coordination system \cite{SKANTZE2021101178}. Therefore, humans likely do not distribute their cognitive load equally to track every participant. They structurally reduce the interaction to a primary exchange by prioritizing the current floor holder and the next projected speaker. By mapping group dynamics onto these relative roles, the model can more easily scale to different group sizes. This approach bypasses the combinatorial explosion of joint modeling and allows the system to handle flexible group sizes without retraining. 


The contrast between the Fisher telephone corpus and our AVCC dataset, shown in Table \ref{tab:dynamics}, highlights the structural differences between constrained dyadic speech and unconstrained multiparty interaction. The performance drop from the Fisher corpus to the AVCC dataset highlights the inherent difficulty of in-the-wild interactions. Telephone conversations enforce a strict alternating rhythm because speakers lack visual feedback. In contrast, unconstrained video data have longer gaps and pauses with more open floor states and visual backchannels. This domain shift makes prediction significantly harder. Training on unedited streams ensures the model encounters these natural conversational rhythms, which are artificially truncated by jump cuts in many datasets.

\subsection{Limitations}

While the Role-Relative Projection effectively simplifies the multiparty problem, it inherently introduces a severe class imbalance in the training data. Because speakers are strictly ordered into 'Current' and 'Next' based on activity, states representing a continuation of speech (holds) occur significantly more frequently than states representing a change in speaker (turn shifts). The model also relies on the history bins to designate the current speaker. This dependence creates a delay of up to two seconds before the system recognizes a new floor holder. Additionally, the visual backbone relies on a standard ASD module. This module separates speaker tracks but misses subtle facial expressions. Upgrading to a more detailed visual encoder could improve early turn shift detection. Finally, the architecture supports variable-sized groups, but our evaluation only covers groups of two and three speaking English. Testing on larger groups in diverse languages is required to confirm actual scalability.

\section{Conclusion}

We introduced MuVAP, a multimodal framework for predicting turn-taking in unconstrained multiparty settings. To address the scaling bottlenecks of joint modeling, we proposed a Role-Relative Projection that compresses complex group dynamics into a scalable pairwise state. Crucially, MuVAP operates on a single monaural audio stream and a single camera view, bypassing the traditional requirement for microphone arrays or spatial audio. We demonstrated that visual grounding via ASD tracks can effectively substitute for spatial separation, allowing the model to resolve speaker attribution even during competitive overlaps in a single-channel mix. On the AVCC dataset, MuVAP demonstrates consistent improvements in preemptive turn-taking prediction. Our results suggest that combining scalable state reduction with audiovisual grounding is a robust path for deploying responsive conversational agents in real-world social environments using only standard commodity hardware. Future work will transition this framework from offline evaluation to live deployment. Specifically, we plan to integrate MuVAP into a physical robot architecture to evaluate real-time, closed-loop turn-taking performance with human subjects.

\section{Acknowledgment}
This work was supported by the Wallenberg AI, Autonomous Systems and Software Program (WASP) funded by the Knut and Alice Wallenberg Foundation (KAW), and the Swedish Research Council
project 2020-03812.

The computations and data handling were enabled by the Berzelius resource provided by the Knut and Alice Wallenberg Foundation at the National Supercomputer Centre.

\section{Generative AI Use Disclosure}

Generative AI tools were used in this paper exclusively for editing and polishing the text to remove grammatical errors. These tools were not used for writing any major parts of the paper.

\bibliographystyle{IEEEtran}
\bibliography{mybib}

\end{document}